\documentclass{mhd}     
\usepackage{graphicx}   	

\mhdhead{0}{0}{1}	

\title{Numerical simulations for the DRESDYN precession dynamo}	
\author{A.~Giesecke\inst{1}, T.~Albrecht\inst{2}, G.~Gerbeth\inst{1}, T.~Gundrum\inst{1}, F.~Stefani\inst{1}}
\institute{Helmholtz-Zentrum Dresden-Rossendorf, 
  P.O. Box 510119, D-01314 Dresden, Germany 
  \and Dep. of Mech. and Aerospace Engineering, Monash University, VIC 3800, Australia} 

\begin{document}

\maketitle

\begin{abstract}%
The next generation dynamo experiment currently under development at
Helmholtz-Zentrum Dresden-Rossendorf (HZDR) will consist of a
precessing cylindrical container filled with liquid sodium.  We
perform numerical simulations of kinematic dynamo action applying a
velocity field that is obtained from hydrodynamic models of a
precession driven flow. So far, the resulting magnetic field
growth-rates remain below the dynamo threshold for magnetic Reynolds
numbers up to ${\rm{Rm}}=2000$. 
\end{abstract}

\section{Introduction.}

Planetary magnetic fields are generated by the dynamo effect, the
process that provides for a transfer of kinetic energy from a flow of
a conducting fluid into magnetic energy. Usually, it is assumed that
these flows are driven by thermal and/or chemical convection but
other mechanisms are possible as well. 
In particular, precessional forcing due to (regular) temporal changes
of the orientation of Earth's rotation axis has long been
discussed as a complementary power source for the geodynamo
\cite{malkus, tilgner}.   

The basic principle of a fluid flow driven dynamo has been
successfully demonstrated in three different experimental configurations,
all of which using a more or less artificial flow driving 
\cite{gailitis, stieglitz, monchaux}. 
Further progress is expected from present and future dynamo
experiments like the Madison plasma dynamo experiment (MPDX,
\cite{madison}, the liquid metal spherical couette experiment at the
University of Maryland \cite{maryland} or the planned precession
dynamo experiment that will be designed in the framework of the liquid
sodium facility DRESDYN (DREsden Sodium facility for DYNamo and
thermohydraulic studies, \cite{stefani}).

Precession driven dynamos were found in simulations with a critical
magnetic Reynolds number of the order of 1000 in a sphere
\cite{tilgner}, cylinder \cite{nore}, spheroid \cite{wu}, ellipsoid
\cite{julia}, and cube \cite{krauze}.  Experimentally, a precessing
magnetohydrodynamical flow has been examined by R. Gans \cite{gans2}
in a cylinder with height $H=0.25~\rm{m}$, rotating with $\omega=60
~\rm{Hz}$ and precessing with $\Omega=0.83~\rm{Hz}$ (thus yielding a
precession ratio of $\Gamma=\Omega/\omega\approx 0.0138$).  In these
experiments an externally applied magnetic field was amplified by a
factor of $\sim 3$ but the magnetic Reynolds number of this setup
remains to small to cross the dynamo threshold.

In order to achieve the required large magnetic Reynolds number
indicated by the above listed numerical studies, the scheduled precession
dynamo experiment at HZDR must represent a greatly enlarged version of
this previous setup making the construction of the experiment a challenge. 
\begin{figure}[t!]
\begin{center}
\includegraphics[width=0.675\textwidth]{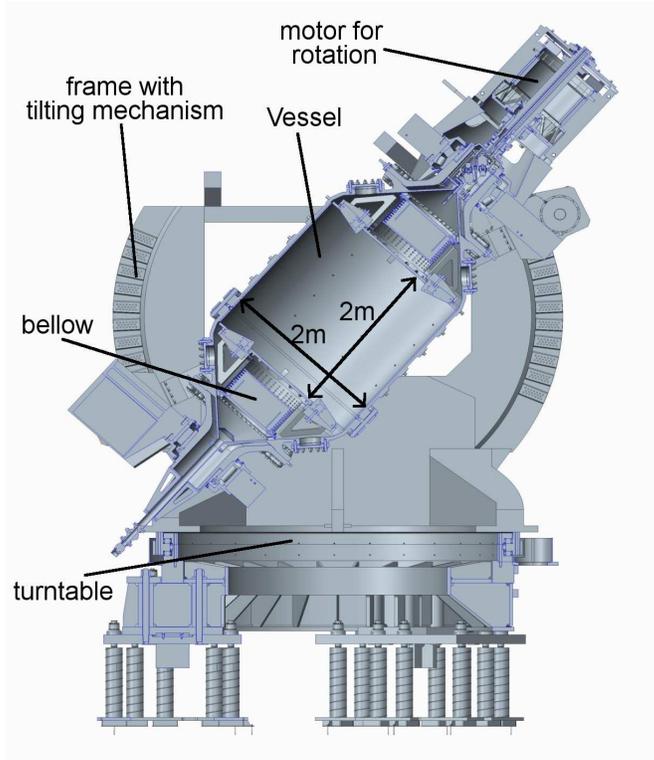}
\caption{Sketch of the planned precession experiment. Diameter and
  height of the cylindrical container will be approximately $2
  ~\rm{m}$. The precession angle can be varied from $\alpha=45^{\circ}$
  to $\alpha=90^{\circ}$ (i.e. $\bvec{\omega}\perp\bvec{\Omega}$).}
\label{fig::exp1}
\end{center}
\end{figure}
The setup will consist of a cylindrical
container with height $H=2\,{\rm{ m}}$ and radius 
$R=1\,{\rm{m}}$ filled with liquid sodium. The cylinder may rotate with a
frequency of up to $\omega=10\,{\rm{Hz}}$ and
precess with up to $\Omega=1\,{\rm{Hz}}$ (see left panel in figure
\ref{fig::exp1}).  
In contrast to previous dynamo experiments no internal blades, propellers
or complex systems of guiding tubes will be used for the optimization
of the flow properties. %

\section{Hydrodynamic flow properties.}

\begin{figure}
\begin{center}
\includegraphics[width=0.75\textwidth]{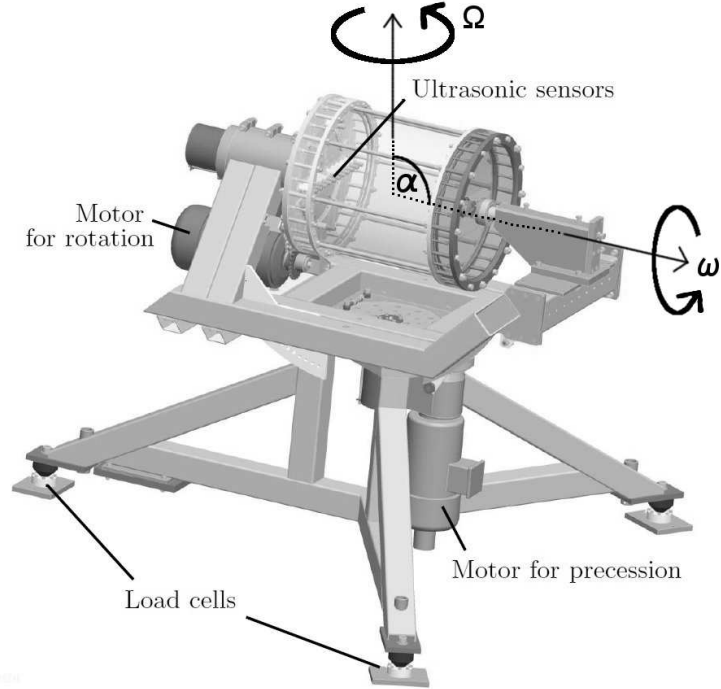}
\caption{Sketch of the water model experiment. The dimensions of the cylinder
  are roughly six times smaller than the planned liquid metal
  experiment.}
\label{fig::exp2}
\end{center}
\end{figure}

A small scale water experiment is in operation in order to investigate
the essential operation parameters for the liquid metal experiment,
such as gyroscopic moments and associated load for the foundation,
motor power consumption, typical flow pattern and flow amplitude
(figure~\ref{fig::exp2}). This experiment is intended to supplement
previous studies that were conducted as part of the French ATER
experiment \cite{leorat1, leorat2, mouhali}.  Most probably, the
precession driven flow will be less suitable for dynamo action than
the flow in highly optimized setups used in the dynamo experiments in
Riga, Karlsruhe or Cadarache.  In order to narrow suitable parameter
regimes that may allow for dynamo action flow properties are estimated
in dependence of precession angle $\alpha$, precession ratio
$\Gamma=\Omega/\omega$ and Reynolds number ${\rm{Re}}=\omega R^2/\nu$.
These properties will be included in kinematic simulations of the
induction equation which are used to estimate the ability of different
flow fields to provide for dynamo action.

In the water experiment axial velocity profiles
at different radial positions were measured using {\it{Ultrasonic Doppler
Velocimetry}} (UDV).  First results confirm observations of
\cite{leorat1, leorat2} that precession provides an efficient flow
forcing mechanism which yields bulk flow speeds of the order of one
fifth to one third of the rotation speed of the container.  In the
liquid metal experiment this will correspond to flow velocities of up
to $20~{\rm{m}}/{\rm{s}}$ so that a rather violent flow is expected.
Based on the rotation speed of the cylinder at $\omega=10~{\rm{Hz}}$,
and the diffusivity of liquid sodium ($\eta=0.08~{\rm{m}}^2/{\rm{s}}$
at $400~{\rm{K}}$), we expect 
magnetic Reynolds numbers of ${\rm{Rm}}=\omega R^2/\eta \sim 750$
which is rather close to the critical 
${\rm{Rm}}$ reported by \cite{nore} from simulations of dynamo action in
a precessing cylinder.

\subsection{Transition to a chaotic state}

\begin{figure}[h!]
\includegraphics[width=0.95\textwidth]{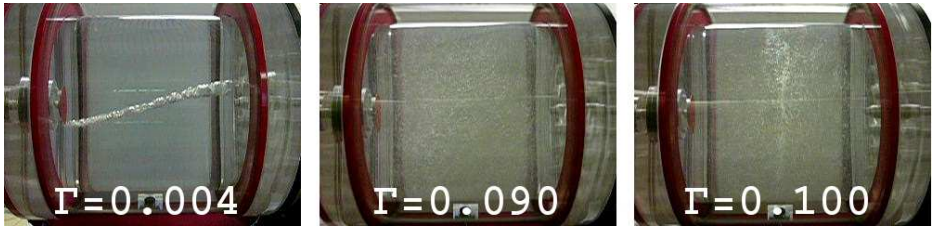}
\caption{Snapshots of the precessing water cylinder at different
  precession rates. The container axis is aligned along the horizontal
  plane and the system precesses around the vertical axis
  ($\alpha=90^{\circ}$). Striking feature is the abrupt transition at
  critical precession ratio $\Gamma^{\rm{crit}}\approx 0.07$ from a
  laminar state (left panel) to more disordered chaotic behavior
  (central panel). For even larger $\Gamma$ the bulk fluid essentially 
is rotating around the precession axis (right
panel).}\label{fig::picprecession} 
\end{figure}

The most striking feature in the water experiment is an abrupt
transition at a critical precession ratio $\Gamma_{\rm{c}}\approx
0.07$ from a laminar state to a disordered chaotic behavior (see
figure~\ref{fig::picprecession}).  The transition goes along with a
sharp increase of the required motor power. The flow properties
change significantly in the chaotic state with the simple $m=1$
Kelvin mode being suppressed, so that we expect a different regime for
the dynamo as well.

\subsection{Comparison with numerical simulations}

For small Reynolds numbers the measurements are compared with
simulations applying the code SEMTEX \cite{blackburn}.  The code uses
a spectral element method and a Fourier decomposition for the
numerical solution of the Navier-Stokes equation which in the
precessing frame reads:
\begin{equation}
\frac{\partial}{\partial
  t}\bvec{u}+(\bvec{u}\cdot\nabla)\bvec{u}+2(\bvec{\omega}+\bvec{\Omega})\times
\bvec{u}=\nu\nabla^2\bvec{u} + \nabla\Phi\label{eq::ns}
\end{equation}
with the boundary condition $\bvec{u}=\bvec{\omega}\times\bvec{r}$.
In equation~(\ref{eq::ns}) $\bvec{u}$ denotes the flow field,
$\bvec{\omega}$ the rotation of the container, $\bvec{\Omega}$ the
precession, $\nu$ the viscosity and $\Phi$ a reduced pressure that
includes centrifugal forces.  Note that equation~(\ref{eq::ns})
describes the precession problem in the so-called turntable frame in
which the observer co-rotates with the precession looking at the
rotating cylinder.

\begin{figure}[h!]
\includegraphics[width=0.975\textwidth]{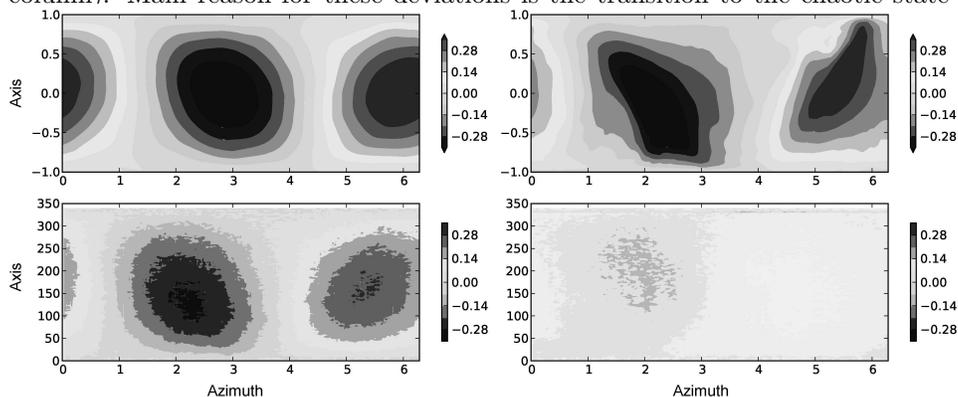}
\caption{Comparison of the axial velocity at $r=0.74$ from
  hydrodynamic simulations (top row) and experimental measurements
  (bottom row). Left: $\Gamma=0.06$, right:
  $\Gamma=0.10$. The experimental data are measured at a rotation
  rate of $\omega=0.2~{\rm{Hz}}$ corresponding to ${\rm{Re}}\sim
  33000$.}\label{fig::compexp}  
\end{figure}

\begin{figure}
\includegraphics[width=0.975\textwidth]{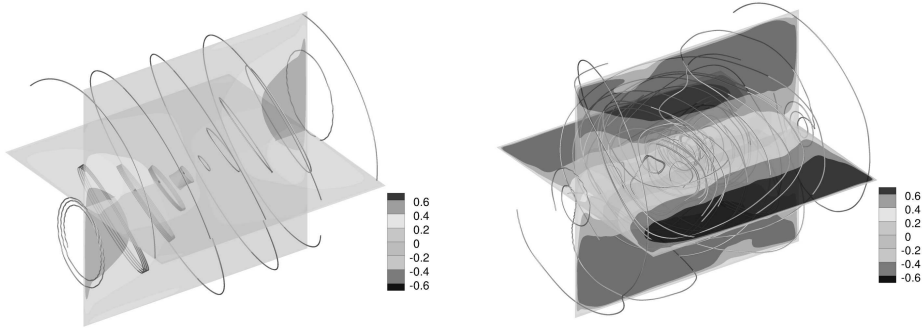}
\caption{Streamlines from simulations with weak precessional forcing
  (left, $\Gamma=0.03$) and strong precessional forcing (right,
  $\Gamma=0.15$). Note the bulk flow rotating around the precession
  axis in the latter case.}\label{fig::streamlines}
\end{figure}

A comparison between numerical solutions and experimental measurements 
is shown in figure~\ref{fig::compexp}. For moderate precession ratio
($\Gamma=0.06$) the pattern and the amplitude of the velocity field
are very similar (left column of figure \ref{fig::compexp}). 
The agreement is worse for a larger precession
ratio ($\Gamma=0.10$) where we observe a larger flow amplitude in
the simulations compared to the experiment (right column). 
Main reason for these deviations is the transition to the chaotic
state in the experiment (around $\Gamma\approx 0.07$) in which the
fundamental $m=1$ Kelvin mode is suppressed.  A vague evidence for
such a transition in the simulations occurs only at significantly
larger $\Gamma\approx 0.15$.  This is indicated in
figure~\ref{fig::streamlines} where the 
rotation axis of the bulk flow has changed its orientation from
roughly parallel to the symmetry axis of the container (left panel) to
a perpendicular direction (aligned with the precession axis, right
panel). 
The change of the orientation of the fluid axis is much less obvious
(abrupt) in the simulation than in the water experiment which may be
due to the much larger ${\rm{Re}}$ in the experiment.

\section{Kinematic simulations of the induction equation.}

In the following we concentrate on the more laminar regime with the 
main fluid rotation axis oriented (more or less) parallel to the
container symmetry axis.
We use different three dimensional velocity fields as an
input for a kinematic solver for the magnetic induction equation which
reads 
\begin{equation}
\frac{\partial}{\partial t}\bvec{B}
=\nabla\times(\bvec{u}\times\bvec{B}-\eta\nabla\times\bvec{B}).\label{eq::ind} 
\end{equation}
In equation~(\ref{eq::ind}) $\bvec{B}$ denotes the magnetic flux
density, $\bvec{u}$ the velocity field and $\eta$ the magnetic
diffusivity.  The numerical solution of equation~(\ref{eq::ind}) is
computed using the numerical scheme presented in \cite{ich}.  The
resulting growth-rates $\gamma$ allow a quick estimation whether a
given flow field is capable to drive a dynamo. The approach works well
in the vicinity of the dynamo onset, but does not allow a
consideration of non-linear effects like the magnetic back-reaction on
the flow.

We use three different patterns for the velocity field. 
The simplest case (case I) makes use of analytic expressions that
describe the fundamental inertial wave with
azimuthal wave number $m=1$ (Kelvin mode). These solutions 
result from the linear
in-viscid approximation of equation~(\ref{eq::ns}) and neglect the
boundary layer flow as well as any non-linear interactions. The
components of $\bvec{u}$ are explicitly given by \cite{manasseh1,manasseh2}
\begin{eqnarray}
u_r \!\!&\!\!=\!\!&\!\!
-\frac{1}{1-\omega_n^2}\!\left[\frac{dJ_1(2\lambda_nr)}{{\rm{d}}r}\!+
\!\omega_n\frac{1}{r}J_1(2\lambda_nr)\right]\!\cos\!\left[k\pi\!\left(\frac{z}{H}\!+
\!\frac{1}{2}\right)\right]\!\sin(\varphi\!+\!t)\nonumber\\  
u_{\varphi} \!\!&\!\!=\!\!&\!\!
-\frac{1}{1-\omega_n^2}\!\left[\omega_n\frac{dJ_1(2\lambda_nr)}{{\rm{d}}r}\!
+\!\frac{1}{r}J_1(2\lambda_nr)\right]\!\cos\!\left[k\pi\!\left(\frac{z}{H}\!
+\!\frac{1}{2}\right)\right]\!\cos(\varphi\!+\!t)\\ 
u_z \!\!&\!\!=\!\! &\!\! 
\frac{k\pi}{h}J_1(2\lambda_nr)\sin\left[k\pi\left(\frac{z}{H}\!
+\!\frac{1}{2}\right)\right]\sin(\varphi+t).\nonumber\label{eq::kelvin} 
\end{eqnarray}
In equation~(\ref{eq::kelvin}) $J_1$ denotes the cylindrical Bessel-function of
first kind, $H$ is the aspect ratio (height over radius), $k$ is the axial
wavenumber, and $t$ is the time scaled by the forcing frequency $\omega$. 
$\lambda_n$ are the radial wave numbers which are computed from the
dispersion relation:
\begin{equation}
\lambda\frac{{\rm{d}}J_1(\lambda)}{{\rm{d}}r}+\omega J_1(\lambda)=0\label{eq::disp}
\end{equation}
(which enforces the radial boundary conditions) and the
eigenfrequencies $\omega_n$ result from the requirements imposed by the axial
boundary conditions with integer axial wavenumber $k$ which leads to
\begin{equation}
\omega^2_n=1+\left(\frac{\lambda_{n}H}{k\pi}\right)^2
\end{equation}
with $\lambda_{n}$ the $n$th root of~(\ref{eq::disp}).  More
elaborate expressions for the linear solutions of~(\ref{eq::ns}) that
include the Ekman pumping and boundary layers are specified in
\cite{zhang}.  However, in this study we are only interested on the
critical magnetic Reynolds number required for the simplest possible
flow pattern that can be excited by a precessional forcing. Hence, we
normalize equations~(\ref{eq::kelvin}) so that $u_{\varphi}(r=R)=1$
and vary the magnetic diffusivity $\eta$ in order to change the
magnetic Reynolds number defined by ${\rm{Rm}}=\omega R^2/\eta$.  In a
second step we might judge if this critical value corresponds to some
reasonable flow amplitude. Any further effects resulting from boundary
layers or higher azimuthal modes (triads) are ignored.

In order to estimate the impact of viscous boundary layers and
non-linear interactions we apply flow fields resulting from the
numerical simulations briefly described in section 2.  We used data
obtained from runs with precession ratio $\Gamma=0.1$ and
${\rm{Re}}=1500$ (case II) and with ${\rm{Re}}=6500$ (case III),
respectively.

\begin{figure}
\includegraphics[width=0.4925\textwidth]{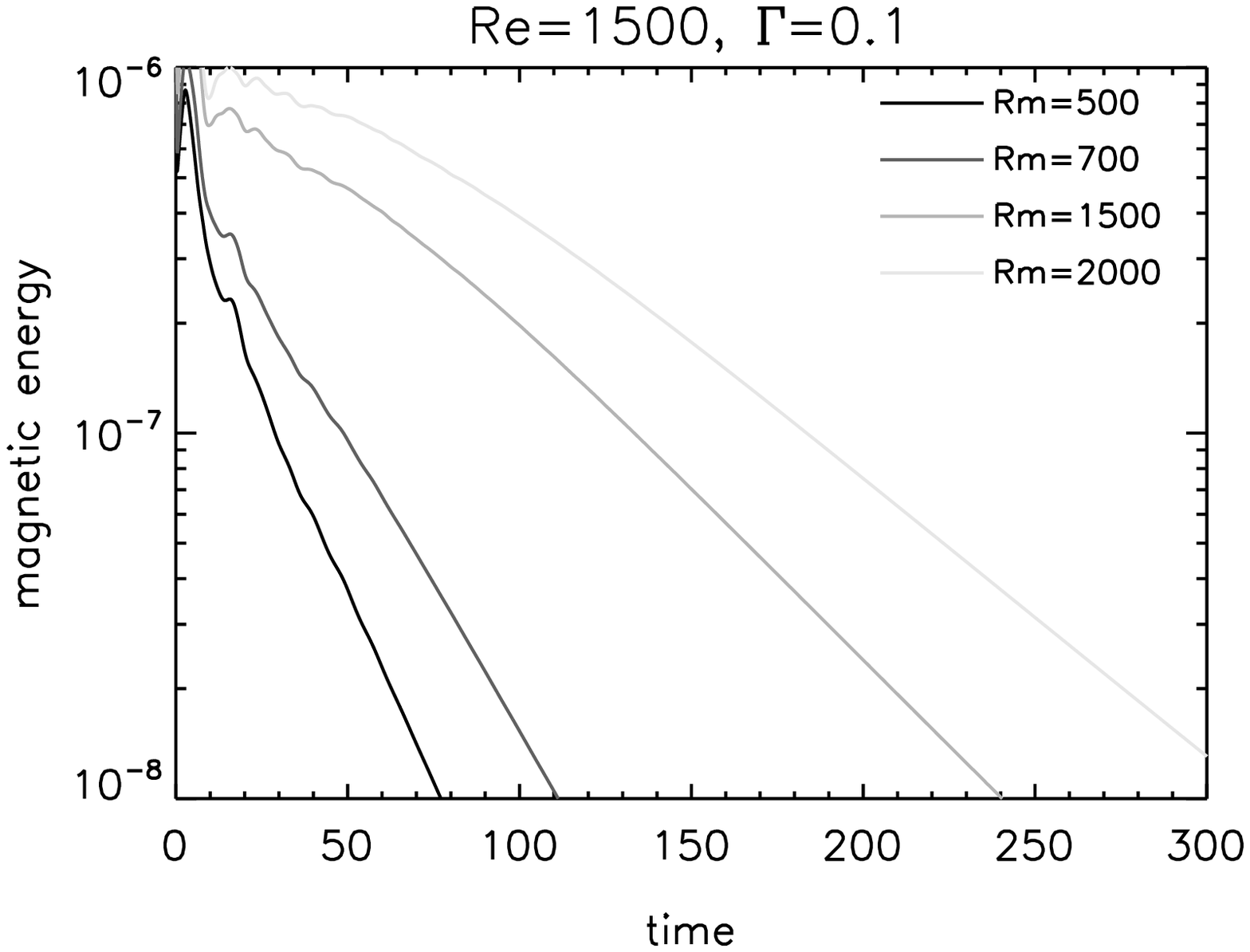}
\includegraphics[width=0.4925\textwidth]{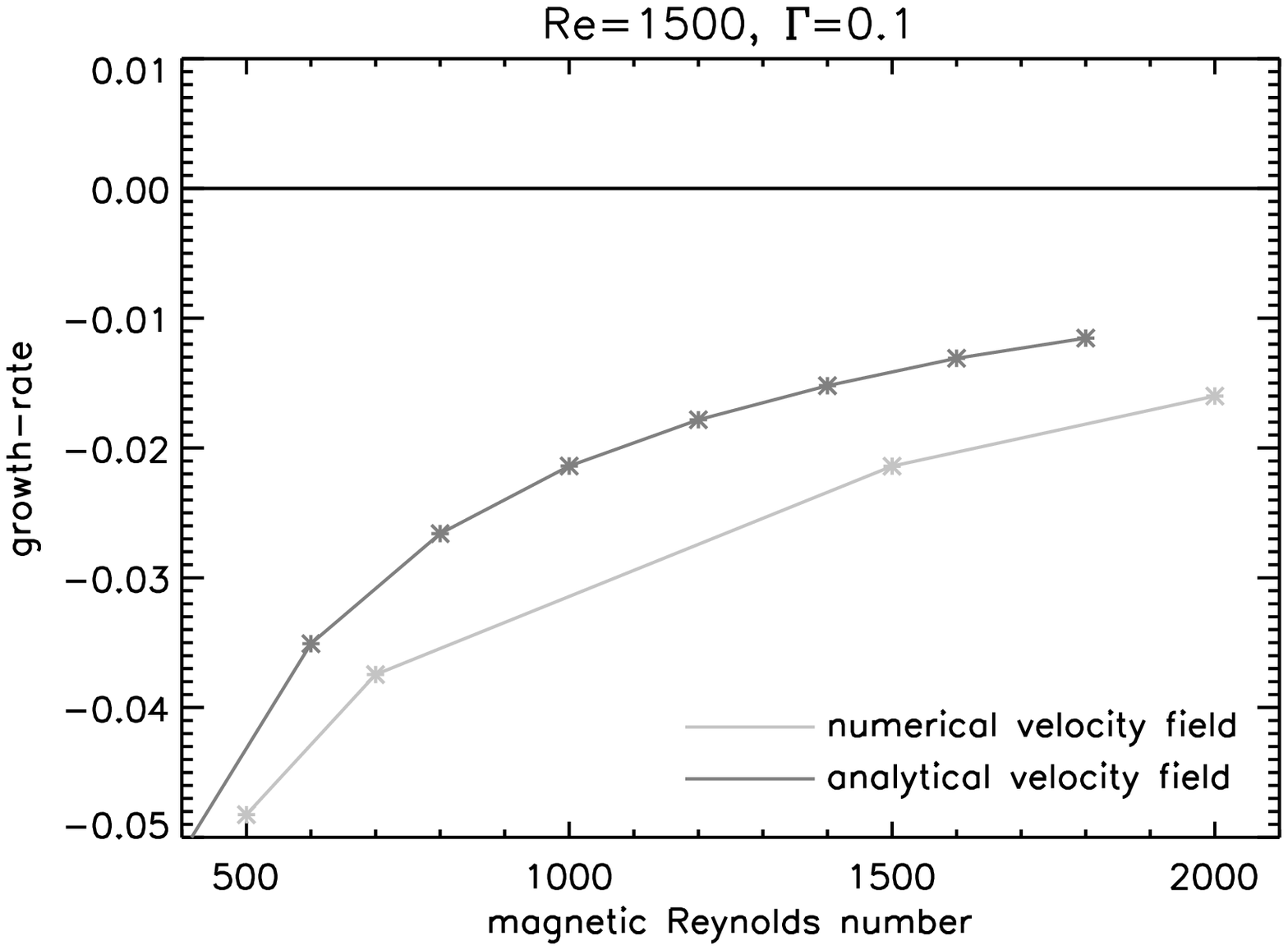}
\caption{Left: temporal behavior of the magnetic energy for a velocity
field obtained at ${\rm{Re}}=1500$ and $\Gamma=0.10$. Right:
corresponding growth-rates of the fundamental dynamo eigenmode versus
${\rm{Rm}}$. }
\label{fig::re1500}
\end{figure}

For ${\rm{Re}}=1500$ the flow is more or less stationary with a rather
simple pattern that is very close to the analytic solutions for the
fundamental $m=1$ Kelvin mode.  Figure~(\ref{fig::re1500}) shows the
temporal behavior of the resulting magnetic energy density
$E_{\rm{mag}}=\frac{1}{2\mu_0}\int\bvec{B}^2dV$ (left panel in
\ref{fig::re1500}).  We do not find any growing solutions up to a
magnetic Reynolds number ${\rm{Rm}}=2000$ and from the behavior of the
corresponding growth rates it seems unlikely that a crossing of the
dynamo threshold occurs within reasonable ${\rm{Rm}}$ (right panel in
Figure~\ref{fig::re1500}).
The dashed curve in the right panel of Figure~\ref{fig::re1500} shows
the growth-rates using the analytic expressions for the Kelvin modes
given in Eq.~(\ref{eq::kelvin}).  The behavior is quite similar to the
growth-rates obtained from the simulations applying the flow field
from the hydrodynamic simulations. Preliminary simulations with even
larger ${\rm{Rm}}$ (not shown) indicate that indeed no crossing of the
dynamo threshold occurs up to ${\rm{Rm}}=5000$.

\begin{figure}[h!]
\begin{center}
\includegraphics[width=0.995\textwidth]{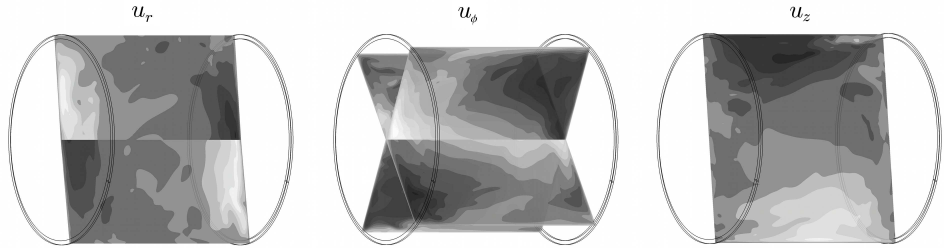}
\end{center}
\caption{Snapshot of the flow components at
  ${\rm{Re}}=6500$. Note that the maximum flow is essentially
  concentrated close to the boundaries.} 
\label{fig::flow_re6500}
\end{figure}
As a next step, we used a flow field obtained from hydrodynamic
simulations at larger Reynolds
number in order to capture the induction effects from a more
time-dependent velocity field. We used a velocity field obtained at
${\rm{Re}}=6500$. At this value the flow is clearly time-dependent but
still dominated by the $m=1$ mode (figure~\ref{fig::flow_re6500}). 

\begin{figure}
\includegraphics[width=0.4925\textwidth]{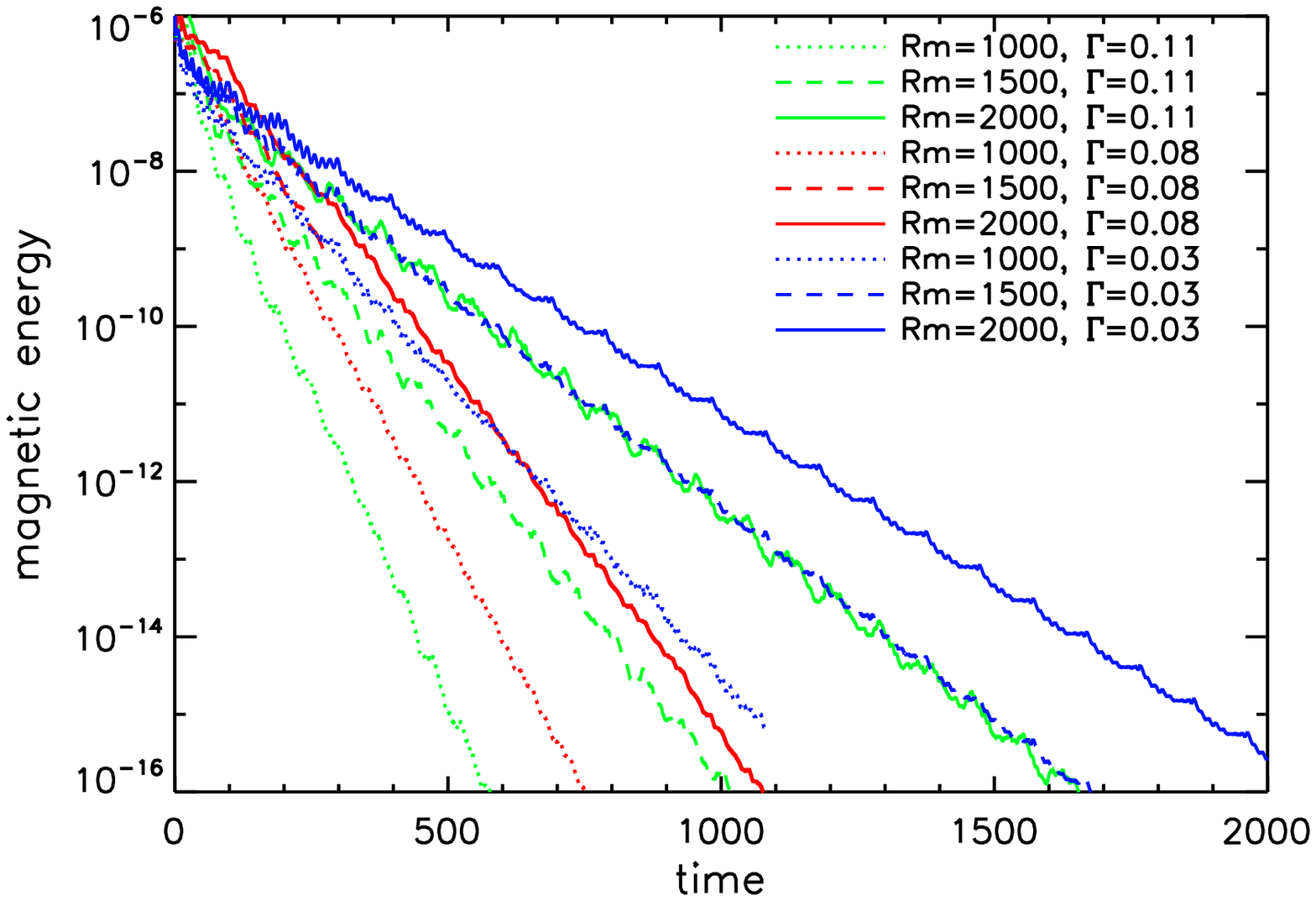}
\includegraphics[width=0.4925\textwidth]{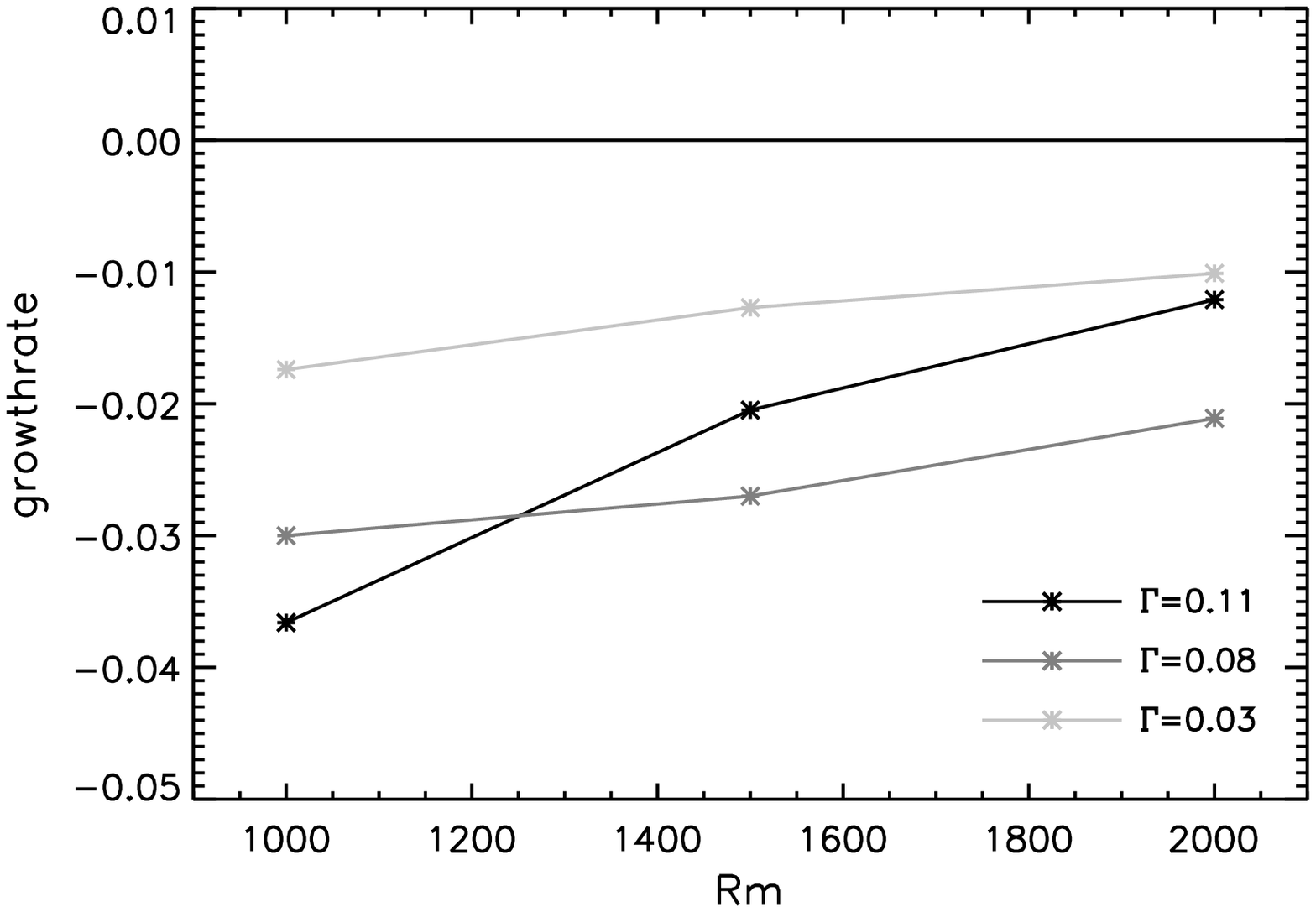}
\caption{Left: temporal behavior of the magnetic field amplitude for a
  velocity field obtained at ${\rm{Re}}=6500$. Right: Corresponding
  growth-rates of the fundamental dynamo eigenmode versus
  ${\rm{Rm}}$.} 
\label{fig::re6500}
\end{figure}

Again, we do not find
any growing solutions for the magnetic field, however, the behavior of
the growth-rates indicates a critical magnetic Reynolds number in the
range of ${\rm{Rm}} =3000...4000$ (figure~\ref{fig::re6500}), which
unfortunately would be far out of reach in the forthcoming dynamo experiment. 
\section{Conclusion.}
So far the kinematic simulations performed within this study did not
show dynamo action. Reasons for this might be the simplistic structure
of the flow which is close to the fundamental Kelvin mode in the low
${\rm{Re}}$ regime or restricted to regions close to the boundaries
for larger ${\rm{Re}}$ (but still dominated by $m=1$). 
The behavior of the growth rates confirms the results from \cite{wietze}
where it was shown that inertial waves cannot drive a dynamo.
However, our simulations were restricted to parameter regimes with
quite low Reynolds number whereas the experiment will be characterized
by ${\rm{Re}}\sim O(10^8)$ and we expect an emergence of more complex
flow structures for more realistic parameters.

Two promising candidates are already known from which we expect an
improvement of the ability of the flow to drive a dynamo. The first
are so-called {\it{triads}} consisting of the forced
fundamental Kelvin mode and two free resonant inertial modes with
larger azimuthal wavenumber. Such triadic resonances have repeatedly
been observed in experiments and simulations of precessing flows
\cite{lorenzani,lagrange}. In a spherical geometry, a subclass of
these modes have a close 
similarity to the columnar convection rolls that are responsible for
dynamo action in geodynamo models and there is little reason to
believe that this should not be the case with a precession driven flow
field. 

\begin{figure}
\hspace*{3cm}
\includegraphics[width=0.5\textwidth]{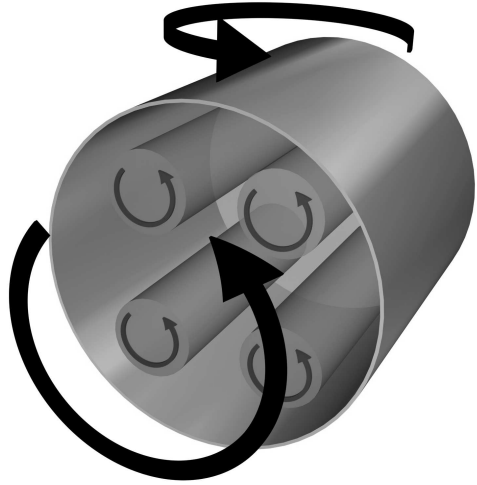}
\caption{Idealized model for the cyclones observed in the ATER
  experiment \cite{mouhali}.}
\label{fig::cyclones}
\end{figure}

The second possibility relies on observations of cyclones in the
French precession experiment ATER \cite{mouhali}.  In that experiment
large scale vortex-like structures emerge for intermediate precession
ratios.  These vortices are oriented along the rotation axis of the
cylindrical container, and, depending on the parameter regime, their
number varies between one and four (figure~\ref{fig::cyclones}).  The
vortices are cyclonic, i.e., their sense of rotation is determined by
the rotation orientation of the cylindrical container.  We suspect
that these vortices provide a significant amount of helicity, but so
far, the axial dependence of their contribution and their interaction
with the fundamental m=1 mode is unknown.  Furthermore, cyclones were
neither observed in the HZDR experiment (so far, no appropriate
velocity measurements in a horizontal plane are available) nor in any
simulations which probably must run at much higher Reynolds number in
order to reveal these modes.

\Thanks{The authors are grateful for support by the Helmholtz
  Allianz LIMTECH and kindly acknowledge discussions with
  C. Nore, J. L{\'e}orat and A. Tilgner.} 




\lastpageno	


\end{document}